\documentclass{PoS}
\def\Dsl{\hbox{/\kern-.6000em D}} 
\def\ltap{\ \raise.3ex\hbox{$<$\kern-.75em\lower1ex\hbox{$\sim$}}\ }
\def\openone{\leavevmode\hbox{\small1\kern-4.2pt\normalsize1}}
\def\Tr{{\rm Tr}}

\title{Large-$N_c$ QCD}

\ShortTitle{Large-$N_c$}

\author{\speaker{Elizabeth JENKINS}
        University of California San Diego\\
        E-mail: \email{ejenkins@ucsd.edu}}

\abstract{A brief review of large-$N_c$ QCD and the $1/N_c$ expansion is given.  Important results for large-$N_c$ mesons and baryons are highlighted.}

\FullConference{International Workshop on Effective Field Theories: from the pion to the upsilon \\
		February 2-6 2009\\
		Valencia, Spain}

\begin{document}

\section{Introduction}

Many features of QCD can be understood by studying $SU(N_c)$ non-Abelian gauge theory in the limit $N_c \rightarrow \infty$~\cite{thooft}.  For $N_c$ large and finite, the expansion parameter away from the large-$N_c$ limit is $1/N_c$.  This nonperturbative expansion parameter is small in QCD where $1/N_c = 1/3$.  

The $1/N_c$ expansion is particularly useful for studying the interactions and properties of large-$N_c$ color-singlet hadrons at low energies.  In this energy regime, gluons and quarks are strongly coupled, but the interactions of  hadrons have an expansion in $1/\sqrt{N_c}$.  Large-$N_c$ mesons are weakly coupled to one another in the 
$1/N_c$ expansion, and become non-interacting and stable in the strict large-$N_c$ limit.  Large-$N_c$ baryons do not decouple from mesons in the large-$N_c$ limit, but their interactions with mesons are determined up to a normalization factor.  Away from the $N_c \rightarrow \infty$ limit, these interactions are modified by subleading 
$1/N_c$ corrections.

The purpose of this talk is to review the formulation of the $1/N_c$ expansion and to highlight important results for QCD mesons and baryons.
For other reviews, see References~\cite{amleshouches,ejannu,ejzac}.

\section{Large-$N_c$ QCD}

An $SU(N_c)$ gauge theory of gluons and quarks
describes the interactions of gluon fields $\left(A^\mu \right)^A$, $A = 1, \cdots, N_c^2 -1$, in the adjoint representation of $SU(N_c)$ and quark fields $q^i$, $i= 1, \cdots , N_c$, in the fundamental representation.  Since there are $O(N_c)$ more gluon degrees of freedom than quark degrees of freedom, gluons dominate the large-$N_c$ QCD dynamics.  At leading order in the $1/N_c$ expansion, `t Hooft showed that quark-gluon Feynman diagrams grow with arbitrarily large powers of $N_c$ unless the limit $N_c \rightarrow \infty$ is taken with $g_s^2 N_c$ held fixed.  This  constraint can be implemented by rescaling the coupling constant $g_s \rightarrow g/\sqrt{N_c}$ in the large-$N_c$ QCD theory.  After this rescaling, Feynman diagrams of given topology scale as 
${N_c}^\chi$, where the Euler character $\chi = 2 -2h -b$, and $h$ and $b$ are the number of handles and (quark loop) boundaries, respectively.  The leading in $N_c$ diagrams, which grow as ${N_c}^2$, are pure glue vacuum diagrams with the topology of a sphere $S^2$ $(h=0,\ b=0)$.  The leading diagrams with a single quark loop $(h=0,\ b=1)$ grow as ${N_c}^1$.  These diagrams can be flattened into a plane bounded by a gluon loop or by the quark loop, respectively, and are called planar diagrams.  Each set of leading diagrams contains an infinite number of diagrams with arbitrary numbers of planar gluon exchanges.  Subleading diagrams contain nonplanar gluon exchange or additional quark loops.    
Each nonplanar gluon exchange in a diagram is suppressed by $1/N_c^2$ relative to the leading planar diagrams, and each additional quark loop is suppressed by $1/N_c$.

The one-loop beta function for the rescaled $SU(N_c)$ gauge coupling $g$ is given by
\begin{equation}
\mu {{d{g}} \over {d\mu}} = -\left( {{11} \over 3} - {2 \over 3} {N_F \over N_c}
\right) {{{g}^3} \over {16 \pi^2}} + O({g}^5).
\end{equation}
Notice that the contribution from diagrams with a quark loop is suppressed by $1/N_c$ relative to the contribution from gluon-loop diagrams.  As the limit $N_c \rightarrow \infty$ is taken, the rescaled coupling $g$ becomes large at a fixed ($N_c$-independent) scale $\Lambda$.
At low energies $E \ltap O\left( \Lambda \right)$, it is assumed that large-$N_c$ QCD becomes confining, so that
the strongly interacting theory of gluons and quarks can be rewritten as an Effective Field Theory (EFT) of color-singlet hadrons.  The expansion parameter of the hadronic interactions in the EFT is $1/\sqrt{N_c}$. 

\section{Large-$N_c$ Mesons}

Large-$N_c$ mesons consist of color-singlet quark-antiquark bound states.  The composite meson operator
\begin{equation}\label{mesonop}
{1 \over \sqrt{N_c}} \ \sum_{i=1}^{N_c} \ \bar q_i \, q^i
\end{equation}
creates a  meson with amplitude of order unity.  The $N_c$-dependence of meson amplitudes is obtained by studying 
the leading $O({N_c}^1)$ planar diagrams with a single quark loop boundary.  The leading $n$-meson amplitude contains $n$ meson operator insertions on the quark loop and is order ${N_c}^{1 - n/2}$.  From this $N_c$-counting, it follows that a meson decay constant is 
$O\left( \sqrt{N_c} \right)$, a meson mass is $O(1)$, a 3-meson coupling vertex is $O(1 /\sqrt{N_c} )$, a 4-meson coupling vertex is $O(1/N_c)$, etc.  A number of other important results follow.  A meson decay width into
two other mesons scales as $\Gamma \sim |{\cal A}_{3-meson}|^2 \sim 1/N_c$, so large-$N_c$ mesons are narrow states which are 
weakly coupled to one another in the nonperturbative expansion parameter $1/\sqrt{N_c}$.  Indeed, in the limit $N_c \rightarrow \infty$, mesons are non-interacting and stable, but for large finite $N_c$, it is important to note that meson widths can become large for highly excited mesons, i.e. with excitations of order $N_c$.  In addition, there are an infinite number of meson states in any given $J^{PC}$ channel.

The effective Lagrangian for large-$N_c$ mesons takes the simple form
\begin{equation}\label{mesonlag}
{\cal L}^{\rm EFT} = N_c \ {\cal L} \left( {\phi \over \sqrt{N_c}} \right),
\end{equation}
where each of the interaction terms in ${\cal L}$ is polynomial in the meson fields $\phi$.     The overall factor of $N_c$ multiplying ${\cal L}$
reflects the ${N_c}^1$ dependence of the leading planar diagrams with a single quark loop.  The $1/\sqrt{N_c}$ accompanying each meson field $\phi$ in $\cal L$ is the normalization factor for the meson operator given in Eq.~(\ref{mesonop}).

There is an extra flavor symmetry for mesons in the $N_c \rightarrow \infty$ limit, sometimes called nonet symmetry.  For finite large $N_c$, the leading diagrams which contain mesons are the $O({N_c}^1)$ planar diagrams with a single quark loop.  Diagrams involving additional quark-antiquark pair creation or annihilation are absent to leading order,  since each additional quark loop is suppressed by $1/N_c$.  
Consequently, there is a
$U(N_F)_q \times U(N_F)_{\bar q}$ flavor symmetry~\cite{veneziano} on the $N_F$ flavors of quarks and antiquarks at leading order which conserves quark number and antiquark number separately.  For $N_F=3$ light flavors, large-$N_c$ mesons transform as a nine-dimensional
$({\bf 3}, {\bf \bar 3})$ representation under the $U(3)_q \times U(3)_{\bar q}$ flavor symmetry.   Under the diagonal subgroup $SU(3)_{q + \bar q}$ (the usual Gell-Mann flavor $SU(3)$ group), this nonet representation breaks to a flavor $SU(3)$ octet and singlet, $({\bf 3}, {\bf \bar 3}) \rightarrow {\bf 8} \oplus {\bf 1}$.  Thus, large-$N_c$ mesons form nonet representations consisting of $SU(3)$ octets and singlets.  For example, the pion octet $\pi$, $K$, $\eta$ and the singlet $\eta^\prime$ form a nonet.  Another important consequence of nonet flavor symmetry is Zweig's rule. 

An example of a meson EFT is provided by the chiral Lagrangian for the pseudoscalar
Goldstone bosons.  The chiral Lagrangian is written in terms of the field $U = e^{2i\Pi/f}$,
where the pion nonet $\Pi = \pi^A T^A + \eta^\prime \openone/\sqrt{6}$.  
The leading $O(p^2)$ chiral Lagrangian is
\begin{equation}
{\cal L}^{(2)} = { {f^2} \over 4} \Tr\ D^\mu U D_\mu U^\dagger +  { {f^2} \over 4} {B}\  \Tr\ \left( {\cal M}^\dagger U + {\cal M} U^\dagger \right),
\end{equation}
where ${\cal M} = {\rm diag} \left( m_u, m_d, m_s \right)$ is the quark mass matrix.  This chiral Lagrangian is of the form Eq.~(\ref{mesonlag}) since the pion decay constant $f \sim O(\sqrt{N_c})$.  The $O(p^4)$ chiral Lagrangian~\cite{gl} is parametrized traditionally in terms of ten low-energy constants (LEC) $L_1 - L_{10}$:
\begin{eqnarray}
{\cal L}^{(4)}
&=& {L_1}\ \left[ {\Tr}\ D_\mu U^\dagger D^\mu U \right]^2
+{L_2}\ {\Tr}\ D_\mu U^\dagger D_\nu U \ {\Tr}\ D^\mu U^\dagger D^\nu U \nonumber\\
&&+{L_3}\ {\Tr}\ D_\mu U^\dagger D^\mu U D_\nu U^\dagger D^\nu U 
+{L_4}\ {\Tr}\ D_\mu U^\dagger D^\mu U \ \ {\Tr} \left( U^\dagger {\cal M} + {\cal M}^\dagger U
\right) \nonumber\\
&&+{L_5}\ {\Tr}\ D_\mu U^\dagger D^\mu U \left( U^\dagger  {\cal M} +  {\cal M}^\dagger U \right)
+{L_6}\ \left[ {\Tr} \left( U^\dagger  {\cal M} +  {\cal M}^\dagger U \right) \right]^2 \nonumber\\
&&+{L_7}\ \left[ {\Tr} \left( U^\dagger  {\cal M} -  {\cal M}^\dagger U \right) \right]^2
+{L_8}\ {\Tr} \left(  {\cal M}^\dagger U  {\cal M}^\dagger U + U^\dagger  {\cal M} U^\dagger  {\cal M} \right)
\nonumber\\
&&-i{L_9}\ {\Tr}\ \left( F^{\mu \nu}_R D_\mu U D_\nu U^\dagger +
F^{\mu \nu}_L D_\mu U^\dagger D_\nu U \right)
+{L_{10}}\ {\Tr}\ U^\dagger F^{\mu \nu}_R U^\dagger F_{L\mu \nu}.
\end{eqnarray}
The $N_c$-dependence of these LECs is naively $O(N_c)$ for terms with a single trace and $O(1)$ for terms with two traces since each additional trace corresponds to an additional quark loop.  It has been shown, however, that
only the linear combination $2 L_1 - L_2$ is $O(1)$, whereas $L_1$ and $L_2$ are each $O(N_c)$.  
The $N_c$-power counting for the LECs is given in Table~1, together with their experimental values extracted at renormalization scale $\mu = m_\rho$~\cite{pich}.  The $N_c$-scaling of the coefficients is evident, although errors are large.  

The LECs can be predicted in terms of the couplings of large-$N_c$ meson resonances to the pion nonet and the masses of the meson resonances.  It has been shown that the LECs are well approximated by the matching obtained from integrating out the lightest $0^{++}$, $0^{-+}$, $1^{--}$ and $1^{++}$ meson nonet resonances~\cite{rxt}.  At $O(N_c)$ in this single-resonance saturation approximation, there are only two independent LECs, which can be taken to be $L_2$ and $L_5$.  The $O(N_c)$ relations are
\begin{eqnarray}
2L_1 &=& L_2, \quad L_9 = 4 L_2, \quad L_{10} = -3 L_2, \nonumber\\
L_3 &=& -3L_2 + {1 \over 2} L_5, \quad L_8 = {3 \over 8} L_5, \quad
L_4 = L_6 = L_7 = 0 \ ,
\end{eqnarray}
which do provide a good fit at $O(N_c)$ to the values quoted in Table~1.

\begin{table}[tbp]\label{table1}
\begin{tabular}{lr@{\hspace{0.2em}}c@{\hspace{0.2em}}lc}
\hline
LEC $\qquad$ & \multispan{3} Value & $1/N_c$ \\
\hline
$2 L_1 - L_2 $ & ($-0.6$ & $\pm$ & $0.6$) $\times 10^{-3}\quad$ & $1$ \\
$L_4$ & ($-0.3$ & $\pm$ & $0.5$) $\times 10^{-3}$ & $1$ \\
$L_6$ & ($-0.2$ & $\pm$ & $0.3$) $\times 10^{-3}$ & $1$ \\
$L_7$ & ($-0.4$ & $\pm$ & $0.2$) $\times 10^{-3}$ & $1$ \\
$L_2$ & ($1.4$ & $\pm$ & $0.3$) $\times 10^{-3}$ & ${N_c}$ \\
$L_3$ & ($-3.5$ & $\pm$ & $1.1$) $\times 10^{-3}$ & ${N_c}$ \\
$L_5$ & ($1.4$ & $\pm$ & $0.5$) $\times 10^{-3}$ & ${N_c}$ \\
$L_8$ & ($0.9$ & $\pm$ & $0.3$) $\times 10^{-3}$ & ${N_c}$ \\
$L_9$ & ($6.9$ & $\pm$ & $0.7$) $\times 10^{-3}$ & ${N_c}$ \\
$L_{10}$ & ($-5.5$ & $\pm$ & $0.7$) $\times 10^{-3}$ & ${N_c}$ \\
\hline
\end{tabular} 
\caption{Experimentally extracted values of the Low Energy Constants of the $O(p^4)$ chiral Lagrangian and their order in the $1/N_c$ expansion~\cite{pich}.}
\end{table}

\section{Large-$N_c$ Baryons}

Large-$N_c$ baryons are color-singlet bound states of $N_c$ valence quarks in a completely antisymmetric
color state~\cite{witten}
\begin{equation}
\epsilon_{i_1 i_2 i_3\cdots i_{N_c}}\ q^{i_1} q^{i_2} q^{i_3} \cdots q^{i_{N_c}} .
\end{equation}
The baryon mass is order $N_c \Lambda$ whereas the baryon size is order
$1/ \Lambda$.  Since the number of quarks inside a baryon grows as $N_c$, but the size of the baryon remains fixed, the density of quarks
inside a baryon becomes infinite in the limit $N_c \rightarrow \infty$.  At large finite $N_c$, the wavefunction of ground state baryons is given by the product of $N_c$ identical quark wavefunctions combined in an $s$-wave.  Since the baryon wavefunction is completely antisymmetric in the color indices of the quarks, it is totally symmetric in the spin-flavor wavefunctions of the quarks. 




The $N_c$-dependence of baryon-meson scattering amplitudes and couplings can be determined by studying quark-gluon diagrams.  An $n$-meson-baryon-antibaryon vertex is order ${N_c}^{1 - n/2}$, which implies that baryon + meson
$\rightarrow$ baryon + meson scattering amplitudes are $O(1)$ in the large-$N_c$ limit, and meson-baryon-antibaryon vertices are 
$O(\sqrt{N_c})$ and {\it grow} with $N_c$.  Consistency of large-$N_c$ power counting rules for baryon-meson scattering amplitudes implies that meson-baryon-antibaryon vertices simplify in the large-$N_c$ limit.  The leading $O(\sqrt{N_c})$ meson-baryon-antibaryon couplings
are determined up to overall normalization~\cite{dm}.  Additional consistency constraints restrict the form of $1/N_c$ corrections to large-$N_c$ baryon matrix elements~\cite{dm,jjj}.    
%

Consider baryon-pion scattering at energies of order unity.  The baryon mass is order $N_c \Lambda$, so the baryon acts as a heavy static source with an $N_c$-independent propagator.  The $B B^\prime \pi$ vertex is order $\sqrt{N_c}$, and
takes the form $\sqrt{N_c} \left( X^{ia} \right)_{B^\prime B} \partial^i \pi^a$.  Here $i=1,2,3$ is a vector spin index and $a=1,2,3$ is a vector isospin index.  
There are two tree diagrams with two $B B^\prime \pi$ vertices contributing to the baryon-pion scattering amplitude at $O(N_c)$, see Fig.~1.  
\begin{figure}
{\includegraphics[width=.4\textwidth]{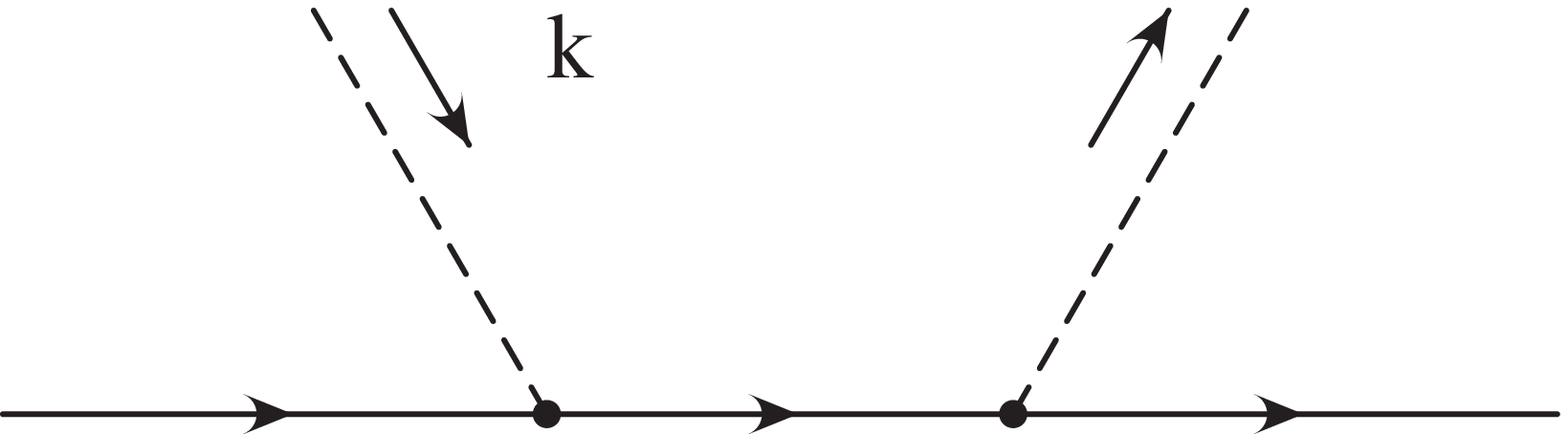}}
\phantom{space}
{\includegraphics[width=.4\textwidth]{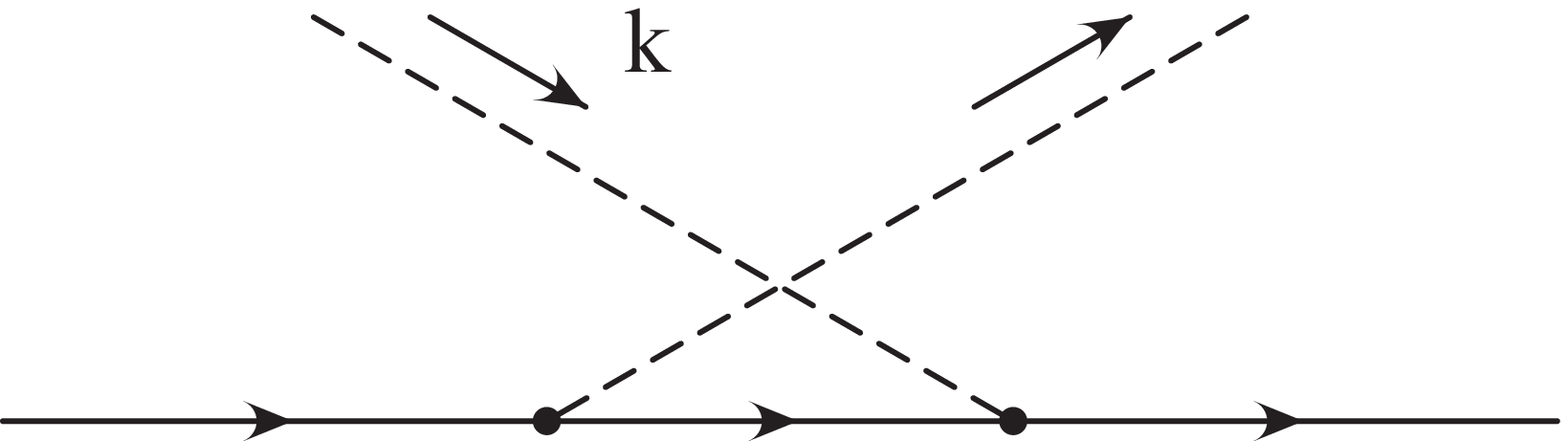}}
\caption{Two tree diagrams contributing to baryon + pion $\rightarrow$ baryon + pion scattering.  Each diagram gives an $O(N_c)$ contribution to the scattering amplitude, but the two $O(N_c)$ contributions must exactly cancel since the overall scattering amplitude is $O(1)$. }
\end{figure}
Although the two diagrams are individually order $N_c$, the sum of the two diagrams is at most order unity by the $N_c$ counting rules, so the baryon-antibaryon-pion matrix elements must satisfy
\begin{equation}\label{dmeq}
{N_c} \left[ X^{ia}, X^{jb} \right] \le O(1) .
\end{equation}
This constraint implies that there is an exact cancellation between the $O(N_c)$ contributions of the two diagrams in Fig.~1.  This cancellation determines all of the $B B^\prime \pi$ matrix elements in terms of a single coupling in the large-$N_c$ limit.  Defining $X_0^{ia} \equiv \lim_{N_c \rightarrow \infty} X^{ia}$, Eq.~(\ref{dmeq}) becomes 
\begin{equation}\label{xcom}
\left[ X_0^{ia}, X_0^{jb} \right] = 0 .
\end{equation}
The $X_0^{ia}$ operators of the large-$N_c$ limit extend the spin and flavor algebras for baryons to a contracted spin-flavor algebra when $N_c\rightarrow \infty$~\cite{dm,gs}.  The usual spin and flavor algebras for baryons are given by  the commutators
\begin{equation}
\left[J^i, J^j \right] = i \epsilon^{ijk} J^k, 
\quad\left[I^a, I^b \right] = i \epsilon^{abc} I^c,
\quad\left[J^i, I^a \right] = 0,
\end{equation}
for $N_F =2$ flavors.
The additional operators $X_0^{ia}$ when $N_c \rightarrow \infty$ commute with one another and transform as vectors under spin and flavor
\begin{equation} 
{\left[J^i, X_0^{ja} \right] = i \epsilon^{ijk} X_0^{ka}, \qquad
\left[I^a, X_0^{ib} \right] = i \epsilon^{abc} X_0^{ic} \ } .
\end{equation}
The above set of commutators defines the exact contracted spin-flavor algebra for large-$N_c$ baryons.
This contracted spin-flavor algebra is related to the usual $SU(4)$ spin-flavor algebra of the quark model
\begin{eqnarray}
&&\left[J^i, J^j \right] = i \epsilon^{ijk} J^k, 
\quad\left[I^a, I^b \right] = i \epsilon^{abc} I^c,
\quad\left[J^i, I^a \right] = 0, \nonumber\\
&&{\left[J^i, G^{ja} \right] = i \epsilon^{ijk} G^{ka}, \qquad
\left[I^a, G^{ib} \right] = i \epsilon^{abc} G^{ic}}, \nonumber\\
&&{\left[G^{ia}, G^{jb} \right] = 
{i \over 4} \delta^{ab}\epsilon^{ijk} J^k +
{i \over 4} \delta^{ij}\epsilon^{abc} I^c\ }, 
\end{eqnarray}
by performing the rescaling ${\lim_{N_c \rightarrow\infty}{ G^{ia} \over N_c} \rightarrow X_0^{ia} }$ of the spin-flavor generators $G^{ia}$.

The spin-flavor generators of the large-$N_c$ baryon spin-flavor symmetry can be used to construct complete and independent operator bases for the $1/N_c$ expansion of static baryon matrix elements.  The operators can be constructed using either $X_0^{ia}$ or $G^{ia}/N_c$ for the spin-flavor generators since these two choices differ at subleading order in $1/N_c$. 
 The form of the operator
expansion is given by~\cite{djm1,cgo,lmr,djm2}
\begin{equation}
 {\cal O}^{\rm m-body}_{\rm QCD} =  {N_c}^m\  
\sum_{n=0}^{N_c}\ \left({1 \over N_c}\right)^n  \  
c_n\ {\cal O}_n ,
\end{equation}
where the operators ${\cal O}_n$ include all independent $n$-body operators in the same spin $\otimes$ flavor representation as the QCD operator. 
Since the operators of the expansion are constructed out of the spin-flavor generators, the matrix elements of the ${\cal O}_n$ are known.  In addition, the order in $1/N_c$ of each operator is known.  The coefficients $c_n$ are unknown, but are $O(1)$ at leading order in $1/N_c$.  These uncalculable coefficients of the $1/N_c$ expansion contain dynamical information of QCD beyond the constraints of large-$N_c$ spin-flavor symmetry and the spin-flavor structure of its breaking in the $1/N_c$ expansion.  Note that for baryons at large finite $N_c$, the $1/N_c$ operator expansion only extends to $N_c$-body operators in the baryon spin-flavor generators.  

Two simple examples for $N_F=2$ flavors suffice to illustrate the usefulness of the $1/N_c$ operator expansion.  For $N_c=3$, the ground state baryons consist of two baryons, the nucleon $N$ and the $\Delta$.  The operator expansion for the baryon mass is 
\begin{equation}
M = m_0 N_c \openone + m_2 {J^2 \over N_c},
\end{equation}
which contains two operators with different orders in $1/N_c$~\cite{jjj}.  These two operators parametrize the two baryon masses.  The $(\Delta -N)$ mass splitting is produced by the $J^2$ operator and is estimated to be a factor of $1/N_c^2$ times the difference of the $\langle J^2 \rangle$ matrix elements of the $\Delta$ and the $N$, or $(15/4 - 3/4)$, times the order $N_c \Lambda$ mass of approximately 1 GeV.  This estimate yields $(\Delta - N) \sim 300~{\rm MeV}$, which implies that the coefficient $m_2$ is order unity as expected.   A second application is to the isovector axial vector currents.
The $1/N_c$ expansion is given by
\begin{equation}
{ A^{ia} = {a G^{ia}} + b {1 \over N_c} J^i I^a + c {1 \over N_c^2} 
\left\{ J^2, G^{ia} \right\}} .
\end{equation}
The three operators in the $1/N_c$ expansion parametrize the three baryon-pion couplings $g_{\pi NN}$, $g_{\pi N \Delta}$ and $g_{\pi \Delta \Delta}$ for the ground state baryons.  Since the matrix elements of $G^{ia}$ are order $N_c$, whereas the matrix elements of $J^i$ and $I^a$ are order unity, the $1/N_c$ expansion can be truncated after the first operator $G^{ia}$ up to corrections of relative order $1/N_c^2$.  This truncation implies that the ratios $g_{\pi N \Delta}/ g_{\pi N N}$
and $g_{\pi \Delta \Delta}/ g_{\pi \Delta N}$ are given by $SU(4)$ spin-flavor symmetry up to corrections of order $1/N_c^2$, or approximately $10 \%$ for QCD~\cite{dm}.  

It is worth emphasizing that the $1/N_c$ expansion yields model-independent predictions which follow only from the spin-flavor structure of the $1/N_c$ expansion.  These predictions are the same in the large-$N_c$ Skyrme Model and large-$N_c$ quark model which both implement the large-$N_c$ baryon spin-flavor symmetry~\cite{am}.  The models, however, often make additional model-dependent predictions, which are not well-satisfied experimentally, as shown in Table~2.  Only the ratio of couplings $g_{\pi N \Delta}/ g_{\pi N N}$ is predicted by the $1/N_c$ expansion.  This model-independent prediction agrees with experiment at the $10\%$ level, as expected.  The Skyrme Model makes the additional prediction of the absolute normalization of the baryon-pion couplings, but this model-dependent prediction is not in good agreement with data.
\begin{table}[tbp]
\label{tab:skyrme}
\begin{tabular}{cccc}
\hline
& Method & Theory & Experiment \\
\hline
$g_{\pi N \Delta}$ & Skyrme Model & $13.2$ & $20.3$ \\ 
$g_{\pi N N}$ & Skyrme Model & $8.9$& $13.5$ \\
$g_{\pi N \Delta}/g_{\pi N N}$ & Large-$N_c$ QCD & $1.5$ & $1.48$\\
\hline
\end{tabular}
\caption{Values for the pion couplings from Adkins, Nappi and~Witten~\cite{anw}. The values of
$g_{\pi N \Delta}$ and $g_{\pi N N}$ in the Skyrme model are model-dependent predictions. The
ratio $g_{\pi N \Delta}/g_{\pi N N}$ obtained in the Skyrme model is model-independent and is the
same as in large-$N_c$ QCD.  The large-$N_c$ QCD prediction for the ratio is corrected at order $1/N_c^2$~\cite{dm}.}
\end{table}

The $1/N_c$ expansion for baryons can be extended to $SU(3)$ flavor symmetry.  The $SU(3)$ flavor symmetry analysis is considerably more complicated than the isospin $N_F=2$ flavor analysis, since $SU(3)$ flavor symmetry breaking is comparable to $1/N_c = 1/3$ and cannot be
neglected.  It is useful to define an $SU(3)$-flavor symmetry breaking parameter $\epsilon \sim m_s/ \Lambda_\chi$.  In the case of baryons containing a single heavy quark $Q=c$ or $b$ in HQET, there is a third expansion parameter describing heavy quark spin-flavor symmetry breaking $\sim \Lambda_{\rm QCD}/m_Q$.

The masses of the ground state baryons for $N_F=3$ flavors provides the most impressive example of $1/N_c$ suppression factors.  The $1/N_c$ expansion of the baryon mass operator is a combined expansion in $1/N_c$ and $SU(3)$ flavor-symmetry breaking $\epsilon$.  The leading $SU(3)$-singlet mass operator is the $0$-body quark operator $N_c \openone$; the most suppressed flavor-${\bf 64}$ operator is third order $\epsilon^3$ in $SU(3)$ flavor-symmetry breaking and suppressed by $1/N_c^3$ relative to the leading singlet operator.  The flavor-singlet, flavor-octet, flavor-$\bf 27$ and flavor-$\bf 64$ mass splittings with definite orders in the $1/N_c$ expansion are listed in Table~3~\cite{jl}, together with their order in flavor-symmetry breaking and $1/N_c$.  These same mass combinations are
plotted in Fig.~2.  The $1/N_c$ suppressions of the various mass combinations are clearly evident in the experimental data.  Moreover, the additional $1/N_c$ suppression factors explain why various $SU(3)$ baryon mass relations work as well as they do.  Flavor-$\bf 27$ mass relations such as the Gell-Mann--Okubo formula and the decuplet equal spacing rule have an extra $1/N_c^2$ suppression factor, and the flavor-$\bf 64$ Okubo mass combination has an additional $1/N_c^3$ suppression factor.  Flavor-octet mass combinations generically have an additional
$1/N_c$ suppression factor.  The $1/N_c$ expansion analysis shows that the three flavor-octet mass combinations split into specific mass combinations suppressed by $1/N_c$, $1/N_c^2$ and $1/N_c^3$, respectively.  The two flavor-$\bf 27$ mass combinations split into specific mass combinations suppressed by $1/N_c^2$ and $1/N_c^3$, respectively. 
\begin{table}
\begin{tabular}{cccc}
\hline
\multicolumn{1}{c}{Mass Splitting}
&\multicolumn{1}{c}{$1/N_c$}
&\multicolumn{1}{c}{Flavor}
&\multicolumn{1}{c}{Expt.} \\
\hline
\smallskip
${{5 \over 8}(2N +3\Sigma
+\Lambda +2\Xi) -{1 \over {10}}(4\Delta +3\Sigma^* +2\Xi^* +\Omega)}$ & $N_c$
& $1$ & * \\
${{1 \over 8}(2N+ 3\Sigma +\Lambda
+2\Xi) -{1 \over {10}}(4\Delta +3\Sigma^* +2\Xi^* +\Omega)}$ & $1/N_c$ & $1$
& ${18.21 \pm 0.03\%}$ \\
${{5 \over 2}(6N -3\Sigma +\Lambda
-4\Xi) -(2\Delta -\Xi^* -\Omega)}$ & $1$ & $\epsilon$ &
${20.21 \pm 0.02\%}$ \\
${{1 \over 4}(N -3\Sigma +\Lambda
+\Xi)}$ & $1/N_c$ & $\epsilon$ & ${5.94 \pm 0.01\%}$ \\
${{1 \over 2}(-2N -9\Sigma
+3\Lambda + 8\Xi) +(2\Delta -\Xi^* -\Omega)}$ & $1/N_c^2$ &
$\epsilon$ & ${1.11 \pm 0.02\%}$ \\
${{5 \over 4}(2N -\Sigma
-3\Lambda +2\Xi) -{1 \over 7}(4\Delta -5\Sigma^* -2\Xi^* +3\Omega)}$ &
$1/N_c$ & $\epsilon^2$ & ${0.37 \pm 0.01\%}$ \\
${{1 \over 2} (2N -\Sigma
-3\Lambda + 2\Xi) -{1 \over 7}(4\Delta -5\Sigma^* -2\Xi^* +3\Omega)}$ &
$1/N_c^2$ & $\epsilon^2$ & ${0.17 \pm 0.02\%}$ \\
${1 \over 4}(\Delta - 3 \Sigma^* + 3
\Xi^* - \Omega)$ & $1/N_c^2$ & $\epsilon^3$ & $0.09 \pm 0.03\%$ \\
\hline
\end{tabular}\caption{$SU(3)$ flavor analysis of ground state baryon masses in the $1/N_c$ expansion. }
\end{table}
\begin{figure}
\centerline{{\includegraphics[width=0.5\textwidth]{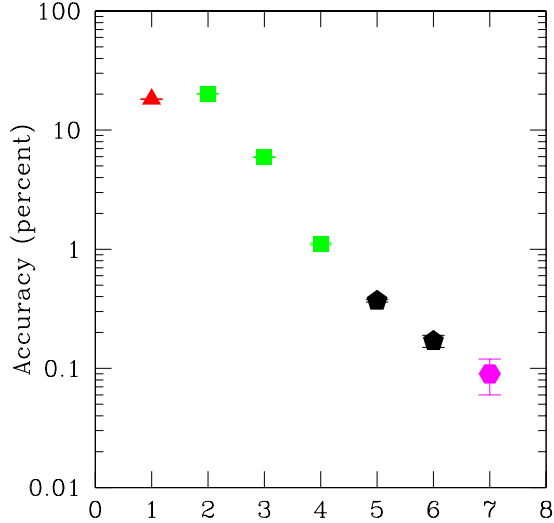}}}
\caption{Experimental accuracies of the mass splittings of the ground state baryons, given in the same order as Table~3.  The accuracies of the mass combinations are order $1/N_c^2$, $\epsilon/N_c$, $\epsilon/N_c^2$, $\epsilon/N_c^3$, $\epsilon^2/N_c^2$, $\epsilon^2/N_c^3$ and $\epsilon^3/N_c^3$.}
\end{figure}

Baryons containing a single heavy quark $Q=c,b$ have a light-quark spin-flavor symmetry in the large-$N_c$ limit
and a heavy-quark spin-flavor symmetry in the heavy quark $m_Q \rightarrow \infty$ limit and in the large-$N_c$ limit.  Heavy-quark spin-flavor symmetry works better for $Qqq$ baryons than for $Q \bar q$ mesons because it also results from the large-$N_c$ limit for baryons.

Many of the masses of baryons containing a single heavy quark $Q$ were predicted based on light quark and heavy quark spin-flavor symmetry in the $1/N_c$ expansion prior to their experimental discovery.  Table~4 gives the mass predictions of Refs.~\cite{jhq1,jhq2} and
the experimental masses discovered afterwards.  The most recently discovered bottom baryon masses were measured one decade after the theory predictions. The $1/N_c$ expansion analysis predicted the central values of the masses, as well as the uncertainty of these predictions,
using mass combinations highly suppressed in the $1/N_c$ expansion and experimentally observed masses as input.  This $1/N_c$ expansion analysis gave more precise predictions than any other theoretical method.  In all cases, the predictions of the $1/N_c$ expansion analysis were successful at the level of stated error bars.  The $N_F=2$ flavor analysis of the heavy baryon mass splittings with strangeness $S=0$, $-1$, and $-2$ are given in Table~5; interesting mass combinations involving both $Qqq$ and $qqq$ baryons are listed in Table~6; and the $N_F=3$ flavor analysis is given in Table~7.   
The $N_Q  m_Q$ operator of Table~6 gives a definition of the heavy quark mass $m_Q$.  The experimental accuracies in Table~7 were computed with the $m_Q$ mass contribution subtracted off in order to produce a dimensionless quantity which can be compared directly to the product of $1/N_c$, $\epsilon$, and $\Lambda/m_Q$ suppressions.  Tables~5-7 are taken from Ref.~\cite{jhq3}.  Once again, there is clear evidence for $1/N_c$ suppression factors in the experimental data.

\begin{table}[tbp]
\begin{tabular}{lr@{\hspace{0.2em}}lc}
\hline
Theory  {($1/N_c$ Expansion)}  & \multispan{2} \qquad Experiment {(Year)} \qquad\qquad    \\
\hline
$\Xi_c^\prime = 2580.8 \pm 2.1\ {\rm MeV}$ & $2576.5 \pm 2.3\ {\rm MeV}$ {(1999)} \\
$\Omega_c^* = 2760.5 \pm 4.9\ {\rm MeV}$ & $2768.3 \pm 3.0\ {\rm MeV}$ {(2006)} \\
$\Xi_b = 5805.7 \pm 8.1\ {\rm MeV}$ & $5774 \pm 11 \pm 15\ {\rm MeV}$ {(2007)} \\
& $5792.9 \pm 2.5 \pm 1.7\ {\rm MeV}$ {(2007)} \\
$\Sigma_b = 5824.2 \pm 9.0\ {\rm MeV}$ & $5811.5 \pm 1.7\ {\rm MeV}$ {(2007)} \\
$\Sigma_b^* = 5840.0 \pm 8.8 \ {\rm MeV}$ & $5832.7 \pm 1.8\ {\rm MeV}$ {(2007)} \\
%
\hline
\end{tabular}
\caption{Theory predictions of Refs.~\cite{jhq1, jhq2} in 1996-7 and subsequent experimental discoveries.}
\end{table}
\begin{table}
\begin{tabular}{cccc}
\hline
\multicolumn{1}{c}{Operator}
&\multicolumn{1}{c}{Mass Splitting}
&\multicolumn{1}{c}{Expt. $Q=c$}
&\multicolumn{1}{c}{Expt. $Q=b$} \\
\hline
{$(m_Q + N_c \Lambda) \openone$} & {$\Lambda_Q$} &{$2286.46 \pm 0.14$}  & {$5620.2 \pm 1.6$} \\
{$J_\ell^2/N_c$} &{$\frac 1 3 \left( \Sigma_Q + 2 \Sigma_Q^* \right)-\Lambda_Q$}  
&{$210.0 \pm 0.5$} & {$205.4 \pm 2.1$} \\
{$J_\ell \cdot J_Q/N_c m_Q$} &{$\Sigma^*_Q - \Sigma_Q$}  
&{$64.4 \pm 0.8$} & {$21.2 \pm 2.5$} \\
\hline
{$(m_Q + N_c \Lambda) \openone$} & {$\Xi_Q$} & {$2469.5 \pm 0.3$} & {$5792.9 \pm 3.0$} \\
{$J_\ell^2/N_c$} & 
{$\frac 1 3 \left( \Xi_Q^\prime + 2 \Xi_Q^* \right)-\Xi_Q$} 
& {$153.7 \pm 0.9$} & \\
{$J_\ell \cdot J_Q/N_c m_Q$} &{$\Xi^*_Q - \Xi^\prime_Q$}   
&{$69.5 \pm 2.3$} & \\
\hline
%
${(m_Q + N_c \Lambda) \openone +}{ J_\ell^2/N_c}$ & $\frac 1 3 \left( \Omega_Q + 2 \Omega_Q^* \right)$ &$2744.7 \pm 2.2$ & \\
{$J_\ell \cdot J_Q/N_c m_Q$} & {$\Omega^*_Q - \Omega_Q $}   
& {$70.8 \pm 1.5$} & \\
\hline 
\end{tabular}
\caption{$Qqq$ baryon masses for baryons with a fixed strangeness $S=0$, $-1$ and $-2$.  The $m_Q$-independence of the $J_\ell^2$ hyperfine mass splittings and the $1/m_Q$ dependence of the $J_\ell \cdot J_Q$ hyperfine mass splittings are seen for the strangeness $S=0$ baryons.}
\end{table}
\begin{table}
\begin{tabular}{cccc}
\hline
\multicolumn{1}{c}{Operator}
&\multicolumn{1}{c}{Mass Splitting}
&\multicolumn{1}{c}{Expt. $Q=c$} 
&\multicolumn{1}{c}{Expt. $Q=b$} \\
\hline
{$N_Q m_Q$} & {$\Lambda_Q - \frac 1 4 \left( 5 N - \Delta \right)$}  
& {$1420.9 \pm 0.2$} & {$4754.6 \pm 1.6$} \\
{$N_Q J_\ell^2/N_c^2 m_Q$} & {$\left[ \frac 1 3 \left( \Sigma_Q + 2 \Sigma_Q^*
\right) - \Lambda_Q \right] -\frac 2 3 \left( \Delta - N \right)$}  
&{$14.4 \pm 0.8 $} & {$9.8 \pm 2.2$}\\
\hline
{$N_Q m_Q$} & {$\Xi_Q 
- \frac 1 4 \left[ \frac 5 4 \left( 3 \Sigma + \Lambda \right) 
- \Sigma^*\right]$}  
& {$1348.4 \pm 0.3$} & {$4671.8 \pm 3.0$} \\ 
{$N_Q J_\ell^2/N_c^2 m_Q$} & {$\left[ \frac 1 3 \left( \Xi^\prime_Q 
+ 2 \Xi_Q^* \right) - \Xi_Q \right] 
- \frac 2 3 \left[ \Sigma^* 
- \frac 1 4 \left( 3\Sigma + \Lambda \right)\right]$}  
& {$13.2 \pm 0.9$}  & \\
\hline
\end{tabular}\caption{Mass splittings of $Qqq$ and $qqq$ baryons.  The $N_Q m_Q$ operator mass splittings, where $N_Q$ is heavy quark number, give a determination of the heavy quark mass $m_Q$.  The $J_\ell^2$ hyperfine mass splittings of $Qqq$ and $qqq$ baryons are simply related up to a correction which is suppressed by $1/N_c^2$ and $\Lambda/m_Q$.}
\end{table}
\begin{table}
\begin{tabular}{cccccc}
\hline
\multicolumn{1}{c}{Mass Splitting}
&\multicolumn{1}{c}{$1/m_Q$}
&\multicolumn{1}{c}{$1/N_c$}
&\multicolumn{1}{c}{Flavor}
&\multicolumn{1}{c}{Expt. $Q=c$}
&\multicolumn{1}{c}{Expt. $Q=b$} \\
\hline
\smallskip
$J_\ell^2$ &$1$ & $1/N_c^2$
& $1$ & ${14.55 \pm 0.04 \% }$ &\\
$T^8$ & $1$ & $1/N_c$ & $\epsilon$
& ${17.22 \pm 0.03 \%}$ & ${16.22 \pm 0.32 \%}$\\
$J_\ell^i G^{i8}$ & $1$ & $1/N_c^2$ & $\epsilon$ &
${3.18 \pm 0.05 \%}$ & \\
$\left\{T^8,T^8 \right\}$ & $1$ & $1/N_c^2$ & $\epsilon^2$ 
& ${0.20 \pm 0.11\%}$ & \\
\hline
$J_\ell \cdot J_Q$
& $1/m_Q$ & $1/N_c^2$ &
$1$ & ${5.36 \pm 0.07 \%}$ & \\
$\left( J_\ell \cdot J_Q \right) T^8$ & $1/m_Q$
&$1/N_c^3$ & $\epsilon$ & ${-0.07 \pm 0.05\%}$ & \\
\hline
\end{tabular}
\caption{Mass hierarchy of baryons with a single heavy quark $Q=c$ or $b$ in a triple expansion in $1/N_c$, $SU(3)$ breaking $\epsilon$, and heavy-quark symmetry breaking $\Lambda/m_Q$.}
\end{table}

The flavor-octet axial vector current couplings of the ground state baryons is another interesting application of the $1/N_c$ expansion.
The octet axial currents for the ground state baryons have the $1/N_c$ expansion
\begin{eqnarray}\label{aia}
A^{ia} = &&{{a_1}} G^{ia} + {{b_2}} {1 \over N_c} J^i T^a 
+ {{b_3}} {1 \over N_c^2} \left\{ J^i, \left\{ J^j, G^{ja} \right\} \right\} 
\nonumber\\
&&+ {{d_3}} {1 \over N_c^2} \left( \left\{ J^2, G^{ia} \right\} - \frac 1 2
\left\{J^i,\left\{ J^j, G^{ja} \right\} \right\} \right)\  . 
\end{eqnarray}
in the $SU(3)$ flavor symmetry limit.
The four coefficients $a_1$, $b_2$, $b_3$ and $d_3$ parametrize the four pion couplings $D$, $F$, $C$ and $H$ of the heavy baryon chiral Lagrangian~\cite{hbchpt}.  The $1/N_c$ expansion in Eq.~(\ref{aia}) can be truncated to the first two operators up to corrections of order $1/N_c^2$, yielding the coupling relations ${C} = -2 {D}$ and ${H} = 3{D}-9{F}$ at this order.  Both of these predictions work very well.  

The pattern of $SU(3)$ flavor symmetry breaking in the flavor-octet axial vector couplings also can be analyzed in the $1/N_c$ expansion.  A new feature of the $1/N_c$ analysis is that the axial couplings obtained from hyperon beta decays are studied together with the strong decay pion couplings of a decuplet baryon to an octet baryon, since both the octet and decuplet baryons are ground state baryons in the same representation of large-$N_c$ spin-flavor symmetry.  Fig.~3 plots the pattern of $SU(3)$ breaking of the baryon flavor-octet axial vector couplings.  There is a clear pattern of $SU(3)$ breaking predicted by the $1/N_c$ expansion of an equal spacing rule linear in strangeness.
This pattern is seen in the decuplet $\rightarrow$ octet pion couplings for baryons with differing strangeness, the first four points in Fig.~3.
The situation for the hyperon beta decay axial couplings is the well-known problem that the data is consistent with no $SU(3)$ breaking.
Thus, a good fit to the data can be obtained, but there is no real explanation for why the $SU(3)$ breaking should be so small.     
\begin{figure}
\centerline{{\includegraphics[width=0.5\textwidth]{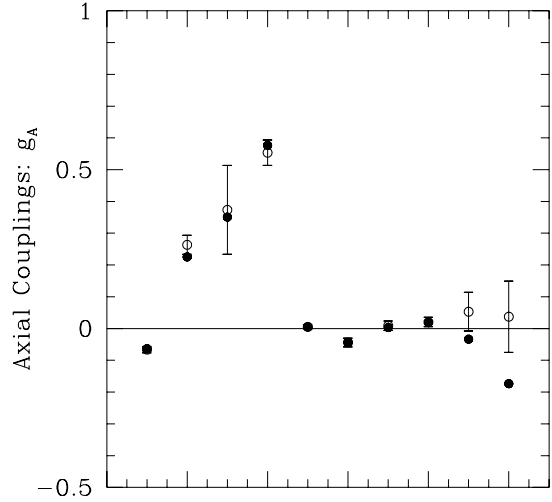}}}
\caption{$SU(3)$ breaking of the axial vector couplings for the ground state baryons.  The $SU(3)$ symmetric fit to the couplings has been subtracted off.  The order of the couplings is $\Delta \to N,\ \Sigma^* \to \Lambda,\ \Sigma^* \to \Sigma,\ \Xi^* \to \Xi,\ n \to p,\ \Sigma \to \Lambda,\ \Lambda \to p,\ \Sigma \to n,\ \Xi \to \Lambda,\ \Xi \to \Sigma$.  The open points with error bars are the experimental data, and the filled points are a fit to the $1/N_c$ symmetry-breaking pattern.  Figure taken from Ref.~\cite{ddjm}.}
\end{figure}

The $U(3)_q \times U(3)_{\bar q}$ flavor symmetry of large-$N_c$ QCD also has important implications for baryons.  At leading order in $1/N_c$, the diagrams involving a large-$N_c$ baryon contain $N_c$ valence quarks with planar gluon exchange and no sea quarks and antiquarks.  Diagrams such as Fig.~4 with an extra quark loop are suppressed by $1/N_c$ relative to the leading diagrams with $N_c$ valence quark lines.
\begin{figure}
\centerline{{\includegraphics[width=1.0\textwidth]{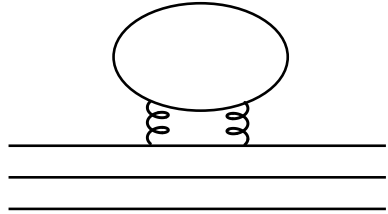}}}
\caption{Baryon diagram which violates $U(3)_q \times U(3)_{\bar q}$ flavor symmetry at relative order $1/N_c$.}
\end{figure}
This planar QCD flavor symmetry implies that baryon flavor octet and singlet amplitudes form flavor nonets at leading order in $1/N_c$~\cite{j}.
For example, the baryon flavor octet axial vector currents $A^{ia}$ and the flavor singlet axial vector current $A^i$ form a nonet at leading
order, which implies that $A^i = A^{i9} + O(1/N_c)$.
The flavor singlet axial current has the $1/N_c$ expansion
\begin{equation}
A^{i} = {{c_1}} J^i + {{c_3}} {1 \over N_c^2} \left\{ J^2, J^i \right\}
\end{equation}
which parametrizes the two $\eta^\prime$ couplings $S_B$ and $S_T$ of the baryon octet and decuplet, respectively.  Truncating the $1/N_c$ expansion of the singlet axial current to the first operator implies the coupling relation $S_T = 3 S_B$ up to a correction of relative order $1/N_c^2$.  The planar QCD flavor symmetry constraint $A^i = A^{i9} + O(1/N_c)$ implies the further coupling relation $S_B = \frac 1 3 (3F -D)$ up to a correction of order $1/N_c$.  Similar remarks hold for the terms of the baryon chiral Lagrangian which are linear in the quark mass matrix $\cal M$.  The three flavor-octet couplings $b_D$, $b_F$ and $c$ satisfy the relation $(b_D + b_F) = -c/3$ to order $1/N_c^2$, whereas the two flavor-singlet couplings $\sigma_B$ and $\sigma_T$ satisfy $\sigma_T = \sigma_B$ to order $1/N_c^2$.  Nonet symmetry relates the flavor-singlet and flavor-octet couplings at leading order in $1/N_c$, implying that $\sigma_B = b_F + O(1/N_c)$.   

\section{Conclusions}

The $1/N_c$ expansion is useful and predictive for QCD hadrons and dynamics.  It is a systematic expansion which yields model-independent results.  The $n$-meson couplings of large-$N_c$ mesons decrease as $(1/\sqrt{N_c})^{n-2}$.  Large-$N_c$ baryons do not decouple from large-$N_c$ mesons in the $N_c \rightarrow \infty$ limit, since the meson-baryon-antibaryon coupling grows as $\sqrt{N_c}$.
In the large-$N_c$ limit, all of the $O(\sqrt{N_c})$ couplings of large-$N_c$ baryons to large-$N_c$ mesons are determined up to normalization by the requirement that exact cancellations occur between the 
$O(N_c)$ contributions to baryon-meson scattering amplitudes.  This cancellation relates all of the ground state baryon pion couplings to one another up to an overall normalization constant.  These coupling relations determine the baryon axial vector flavor-octet matrix elements $X_0^{ia}$ in the large-$N_c$ limit.  This additional $X_0^{ia}$ operator extends the baryon spin and flavor algebras to a contracted spin-flavor algebra in the large-$N_c$ limit.  The consequences of contracted spin-flavor symmetry for static baryon quantities and the spin $\times$ flavor structure of $1/N_c$-suppressed breakings of spin-flavor symmetry have been determined.  There is ample evidence for the $1/N_c$ hierarchy and spin-flavor pattern in the baryon sector.  The pattern of spin-flavor symmetry breaking is particularly intricate since $SU(3)$ flavor breaking is comparable to $1/N_c = 1/3$ for QCD.  The $1/N_c$ expansion has yielded numerous insights for the interactions and properties of hadrons and holds much promise for the future.

\end{document}